\documentclass[preprint,aps,pre,preprintnumbers,amsmath,amssymb,nofootinbib,floatfix]{revtex4-2}

\usepackage{graphicx,color}
\usepackage{url}
\usepackage{textcase}
\usepackage{bm}
\usepackage{tabularx}
\usepackage[T1]{fontenc}

\newcommand{\half}{\frac{1}{2}}
\newcommand{\be}{\begin{equation}}
\newcommand{\ee}{\end{equation}}
\newcommand{\bea}{\begin{eqnarray}}
\newcommand{\eea}{\end{eqnarray}}
\newcommand{\ba} {\begin{align} }
\newcommand{\ea} {\end{align} }

\begin{document}

\title{Scaling Perspectives of Underscreening in Concentrated Electrolyte Solutions}

\author{ Samuel A. Safran$^1$}
\email{sam.safran@weizmann.ac.il}, 
\author{ Philip A. Pincus$^2$}
\email{fyl@ucsb.edu}
\affiliation{$^1$Department of Chemical and Biological Physics\\ 
Weizmann Institute of Science, Rehovot 76100, Israel\\
$^2$Physics and Materials Departments\\ 
University of California, Santa Barbara, CA 93106, United States\\}

\begin{abstract}
We present a scaling view of underscreening observed in salt solutions in the range of concentrations greater than about 1M, in which the screening length \emph{increases} with concentration. The system consists of hydrated clusters of  positive and negative ions with a single unpaired ion as suggested by recent simulations. The environment of this ion is more hydrated than average which leads to a self-similar situation in which the size of this environment scales with the screening length.  The prefactor involves the local dielectric constant and the cluster density.  The scaling arguments as well as the cluster model lead to scaling of the screening length with the ion concentration, in agreement with observations.
\end{abstract}

\maketitle

\def\half{\frac{1}{2}}

\section{Introduction}
\label{paperintro}
In thermodynamic equilibrium, the electric field associated with fixed charges (e.g., on bounding surfaces) may be screened by other mobile charges which may reorganize to optimize the electrostatic free energy. The resulting electric fields are determined in terms of the full charge distribution by the Poisson equation \cite{Safran2018, Andelman1995}. However, this requires knowledge of that same charge distribution which must be self-consistently obtained from the forces acting upon the charges including electrostatics and other interactions, for example excluded volume (hard-core) interactions and the resulting correlations \cite{Kjellander2019, RN27} in the presence of electrostatics. This results in the canonical many body problem. For the case of aqueous solutions of dissolved salts, e.g., NaCl, the Debye-H\"{u}ckel (DH) theory is a long-established model to solve this system in the limit of low concentrations of salt in the aqueous solvent \cite{Levin2004}. This theory relies on several approximations:  (i) The ions rearrange in response to the local, average potential. (ii) The ions are treated as point charges at low concentrations of dissolved salts. (iii) The solvent structure is not taken into account in this primitive model‚ which treats the water as a dielectric continuum (dielectric constant $\epsilon \approx 80$). For a coordinate system with one ion fixed at the origin that defines the radial coordinate r, the DH theory for a 1-1 electrolyte yields a dimensionless electrical potential relative to infinity and temperature, 
\begin{equation}
	\phi = \frac{\ell}{r} \, e^{-\kappa r}.
	\label{potential}
\end{equation}

The DH screening length $\lambda$, is given by

\begin{equation}
	\lambda^{-1}\equiv \kappa_D=\sqrt{8 \pi c \ell}
	\label{kappaD}
\end{equation}

The temperature, $T$, is given in units of Boltzmann's constant; the Bjerrum length, $\ell\equiv e^2/(\epsilon T)$, is the distance between two charges such that their electrostatic energy is $T$; $e$ is the absolute value of the electronic charge, and $c$ is the concentration of dissolved salt molecules.  For water, the Bjerrum length is $\ell\approx 0.7$~nm and the screening length for 1mM salt is $\lambda\equiv \kappa_D^{-1}\approx 10$~nm.  One way to view the physics of this screening is to consider a snapshot of an instantaneous fluctuation of  a local region in the system as  a sphere surrounding a charge fixed at the origin. [We note that this applies to the neighborhood of a charge density fluctuation and is not a mean-field description of the entire system.] The central charge repels co-ions and attracts counter-ions. A sphere that has a radius that exceeds the screening length will contain about one extra counter-ion which then strongly reduces the electric field of the fixed charge via Gauss' Law. Note that if the screening length is smaller than the mean distance between charges, then the continuum approximation for the charge density, appropriate at low concentration, is no longer valid and one expects corrections to the DH screening length, Eq.~\ref{kappaD}. This already occurs in the range of a few millimolar concentrations \cite{Israelachvili1978, Israelachvili1978a}.

Although the DH theory is semi-quantitatively in agreement with a diverse set of observations, direct measurements of the screening length are rare. One approach uses the surface force apparatus (SFA) which measures the force between two parallel surfaces immersed in a fluid \cite{Israelachvili2010, Lin2021}. Typically, the surfaces are mica which is a clay that becomes charged in contact with water by the dissolution of potassium ions and leaving a highly anionic surface. The SFA is well-suited to the aqueous saline system because the interplanar separation can be manipulated in the nanometer range. Then, for two identical parallel surfaces, the variation of the electrostatic repulsive force with distance is exponential which, assuming linear response of the charge distribution to the fixed, surface charges, yields the screening length, at least, in the low concentration range \cite{Raviv2001}.  Already in the 70's, Israelachvili and co-workers investigated aqueous salt solutions using this method \cite{Israelachvili1978, Israelachvili1978a}. Their general conclusions for 1-1 salts is that the DH screening (Eq.~\ref{kappaD}) worked well in the millimolar range but already for 10's of millimolar concentrations there were 20\% discrepancies from Eq.~\ref{kappaD} in the direction of longer screening lengths than predicted. This weakening of the concentration dependence of the screening is called underscreening  and may not be unexpected  given our previous remarks.

More recently, SFA experiments have been extended to higher concentrations (well into the molar range) for both saline aqueous solutions and solutions of ionic liquids \cite{Gebbie2015, RN2, RN86}. In both these situations, anomalous underscreening is observed where there is a minimum in the screening length and, at concentrations in excess of about 1M, there is an increase in the screening length with increasing electrolyte concentration until insolubility. In the neighborhood of the minimum screening length (typically in the nanometer range), there are Kirkwood oscillations \cite{RN145} in the interplanar force that are associated with the correlations that develop in the electrostatic interactions of finite-sized ions with hard-core repulsions. In the presence of such oscillations, the decay length is measured in the distal regime where the oscillations have decayed and there is relatively clean exponential behavior. In this distal, anomalous regime (decay length increasing with salt concentration), Perkin and co-workers \cite{RN86} have shown that the data from different systems (salts dissolved in water and various ionic liquid solutions), collapse onto the scaling relation, written in terms of the inverse experimental decay length,  
\begin{equation}
	\frac{\kappa}{\kappa_D} \sim \frac{1}{(\kappa_D d)^3}
	\label{susan1}
\end{equation}
where $d$ is an appropriate ionic radius. Anomalous underscreening, as embodied in Eq.~\ref{susan1} is controversial. For example, atomic force studies \cite{RN143} through saline solutions against both silica and mica substrates are reported to not exhibit exponential decaying repulsive forces but rather power laws consistent with dispersion (van der Waals) forces. On the other hand, there exist multiple examples of situations in the presence of aqueous salt solutions that exhibit reentrant \cite{Yuan2022} behavior in molar range ionic strengths which may be signatures of anomalous behavior. Several simulation investigations also do not show explicit anomalous underscreening \cite{RN63, RN144} or smaller than expected effects \cite{RN115}. However, recent restricted primitive model (no hydration) simulations \cite{RN59}  do observe concentration fluctuations \cite{RN27} which may be interpreted in terms of weakly charged (nearly neutral) nanoscale amorphous salt clusters, which can lead to weak 
screening, consistent with anomalous underscreening. The goal of this communication is to present a scaling picture for Eq.~\ref{susan1} which has been, in part, stimulated by the H\"{a}rtel et al simulations \cite{RN59}.  The resulting cluster geometries and topologies in the aqueous solution can be very complex \cite{Deviri2020}.  As is often typical of such highly disordered and possibly fractal systems \cite{Lubensky1979}
we proceed with some simplifications which require rather detailed complex renormalization group calculations for justification.  This is outside the scope of this paper and will not be attempted here.  However, we note that Eq.~\ref{susan1} predicts that the screening length,$1/\kappa$ is proportional to the average ion concentration $c$.   This suggests that the system behaves in a self-similar manner, i.e.,  as the ion concentration is increased, the electrostatic behavior is the same if one just rescales the screening length.

\section{Scaling relations}
\label{paperscaling}
Note that the Perkin scaling, Eq.~\ref{susan1}, contains a new length scale, the ionic radius $d$, which is absent in the DH theory since it assumes point charges. The ionic dimension suggests that correlations induced by short range excluded volume interactions could be playing a central role at elevated concentrations \cite{RN142, RN28}.  Indeed, this, along with the Coulomb attractions of oppositely charged ions, is at the heart of the cluster formation \cite{RN59}. This idea is embodied in the following scaling argument that mimics Perkin's experimental finding. We express the inverse screening length as

\begin{equation}
	\kappa=\kappa_D \ \psi(\kappa_D d)
	\label{scaling1}
\end{equation} 
where $\psi(x)$is a scaling function such that $\psi(0)=1$, $\psi(x)$ is assumed to be a power law,
$\psi(x) \sim x^p$  for $x \gg 1$. Guessing that $\kappa$ can also be written as a function of  the dimensionless quantity $cd^3$ in this regime, and writing $\kappa \sim (cd^3)^r$ we determine the exponents $p=-3$ and $r=-1$ by equating powers of $c$ and $d$.  This immediately give scaling that agrees with the experimental Perkin relation. 

Another scaling view that is suggested by the observation of clusters in the simulations by H\"{a}rtel et al. \cite{RN59}, is to posit that the inverse screening length is the DH form, Eq.~\ref{kappaD}, with an effective concentration, $c^{*}$ of charges. The simulations suggest that at high salt concentrations, the ions are organized in nearly neutral clusters with a sub-set of the clusters having one elementary net charge equally likely (in the primitive model) to be positive or negative. The densification of the ions in mainly neutral clusters is already suggested by DH theory where the ``ideal solution" phase of ions in water is unstable at a critical concentration proportional to the inverse Bjerrum volume. When excluded volume effects are considered, the critical point is characterized by a concentration proportional to the inverse ionic volume and a critical value of the Bjerrum length proportional to the ionic size  \cite{RN54}. The clusters are hydrated and possibly fractal \cite{Deviri2020, Lubensky1979} (see below), surrounded by a region whose concentration (of paired ion clusters and water) is different from the average of the surroundings.  We define the effective charge density, $c^{*}$, to be the concentration  of unpaired ions.  Each unpaired ion is associated with a neutral cluster of $Z$ ions where $Z=\rho (4\pi/3) R^3$ and $\rho$ is the ion density in the cluster, see Fig.~\ref{fig:fig1}.  In addition, there are also neutral clusters with no associated unpaired ions and self-similarity suggests that their average size and cluster concentrations are similar to those with one unpaired charge.  Conservation of ions in both these types of clusters means that $c \sim n Z$ where $n$ is the cluster concentration.   The charge density is one unpaired charge per charged cluster and is thus the cluster concentration times unity. This implies that $c^{*} = c/ Z$.  This allows us to smoothly interpolate between the DH limit where there are no clusters and $Z=\rho \, 4 \pi  R^3/3\rightarrow 1$ and the high concentration limit where $Z$ might be large. 

In the spirit of critical phenomena, the extent of the ionic domain $R$, may be associated with the screening length itself, $R \sim \kappa^{-1}$. With this renormalization, the effective charge density is very small and the DH expression for screening with $c^{*}$,  can be used instead of $c$ in Eq.~\ref{kappaD}.   In the next section, we demonstrate that while $R \sim \kappa^{-1}$, it is numerically smaller than the screening length due to the presence of an additional small parameter in the system, the variation in the local dielectric constant.   This self-consistently then results in the scaling form, Eq.~\ref{susan1} with the screening length being proportional to the ion concentration:

\begin{equation}
	\lambda \equiv \kappa^{-1} \sim \big( \frac{\rho}{c \ell \kappa^3} \big)^{1/2} \ \  {\rm{or}} \ \ \lambda \sim c \ell/\rho	\label{renorm}
\end{equation}   


\subsection{Physical picture}
\label{paperphysics}
The remaining paragraphs are devoted to providing a physical picture that relates these scaling views to Coulomb interactions of unpaired ions in the presence of neutral, aqueous clusters of oppositely charged ions, which allows one to estimate the physical quantities that enter Eq.~\ref{renorm}.  Most of the ions in the system are in these neutral clusters (that are surrounded by water, in a possibly complex and fractal geometry).  However, some ions, most probably at the corners of these clusters where there is maximal exposure to the high dielectric constant of the surrounding water (which reduces the local electrostatic energy), can be relatively unpaired, resulting in a net cluster charge. Indeed, studies of finite salt crystals (in vacuum) suggest the importance of corner charges \cite{RN34} in those systems.  The H\"{a}rtel et al. simulations for NaCl in water show that the net charge is most typically due to one unpaired ion in the cluster.   

\begin{figure}
	\centering
	\includegraphics[width=1.2\linewidth]{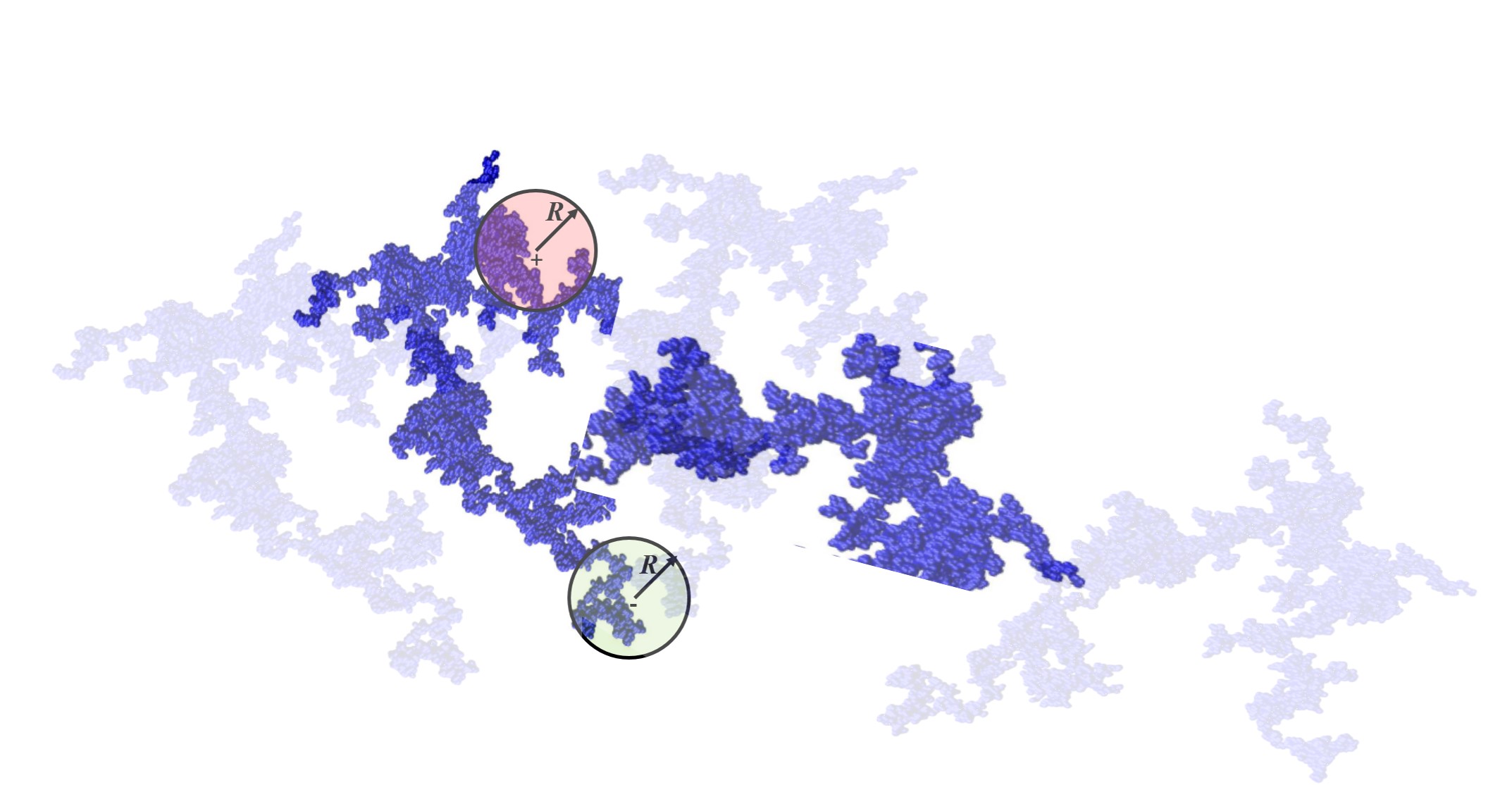}
	\caption{Sketch of clusters with two examples of unpaired ions.  The background clusters are shown in shadow. }
	\label{fig:fig1}
\end{figure}

The argument in the previous section posited that the cluster dimension scales with the screening length.  How might this occur? Assuming that a particular cluster has an associated unpaired ion, that ion must be situated in a location (such as the cluster corner) where the local dielectric constant exceeds the spatially averaged dielectric constant, because then the Coulomb energy associated with that ion will be reduced. The solution may be viewed as a dispersion of such monovalent clusters and as suggested above, these clusters behave much as the point ions in DH theory, except that they have a dimension $R$ as shown in Fig.~\ref{fig:fig1}. The region around the net, single ion, charge of the cluster may be modelled as  a sphere with a reduced, local Bjerrum length (locally higher dielectric constant than its system-wide and concentration dependent average \cite{RN13}) and a potential arising from the  Yukawa-like screening by other clusters at distances exceeding $R$. In this view, one calculates the local Coulomb self-energy, $u_s$, of the charge with its own Coulomb potential (reduced by the locally higher Bjerrum length) as well as its surrounding screening potential in a sphere of radius $R$.  Such a calculation of the self- energy (in units of $T$) of the unpaired ion gives,
\begin{equation}
	u_s=\half \, \left( -\frac{\kappa\,\ell}{1+\kappa R} + \frac{\Delta \ell}{R} \right)
	\label{selfenergy}
\end{equation}
Here, $\Delta \ell>0$, is the difference of the Bjerrum lengths outside and inside the sphere.  Within the sphere, the locally higher dielectric constant (which is the region where a unpaired, charged ion would reside) results in a locally decreased Bjerrum length. For small local changes in the dielectric constant, minimization of this self-energy with respect to the cluster dimension gives $\kappa R \approx (\Delta\ell/\ell)^{1/2}$. In other words, the cluster dimension scales with the screening length but is smaller than it by the variation in the local dielectric constant: the small quantity $\delta^{1/2}$ where $\delta\equiv \Delta \ell/\ell$. As shown in the previous section, the resulting screening length then indeed obeys Perkin scaling but is amplified by a factor of $1/\delta^{3/2}$.  We expect the density of paired ions within the cluster to scale as $\rho \sim 1/d^3$ but with a proportionality factor $\eta$ that may be significantly less than unity if the clusters are very hydrated. For a close-packed crystal $\eta=1$.  We finally can write:
\begin{equation}
	\lambda = \kappa^{-1}=\frac{6\eta \ cd^3 \ell}{\delta^{3/2}} 
	\label{finallambda}
\end{equation}
The parameter $\delta=\Delta \ell/\ell$ and the cluster density $\rho=\eta/d^3$   are rather model sensitive and require independent studies for each model. It is important to recognize that effective ionic size, $d$, and $\delta$ are intertwined and not completely independent for the different water models. In other words, $d^* = d \delta^{-1/2} > d$
may be considered as a new effective ionic size. In hydration models, $d^*$  may
be considered as the hydrated ionic dimension.

\subsection{Caveats}
\label{paperphysicscaveats}

We have presented two scaling arguments that predict that at relatively high salt concentration, the screening length scales linearly with the salt concentration and the Bjerrum length, in agreement with experiments using surface force measurements. Both arguments treat the anions and cations in the same manner even though it is known that for NaCl, the Cl ion is more strongly hydrated that the Na.  Thus, corrections that distinguish the two ions are absent from our picture \cite{Sellner2013}.  The first scaling argument is a purely dimensional one and uses the fact that there are two dimensionless quantities in the system, $\kappa_D d$ and $cd^3$.  The second scaling argument notes, in agreement with the simulations of H\"{a}rtel et al., that the system consists of nearly neutral clusters of local density $\rho$ that accounts for their hydration.  Screening arises from the relatively few unpaired ions in the system with $c^{*}=c/(\rho \, R^3)$. In the spirit of scaling in self-similar systems, if one assumes that $R$ is proportional to the screening length, and then uses DH theory for the screening length as a function of the very dilute concentration of {\it{unpaired}} ions, one obtains $\lambda \sim \ell \, cd^3$.  

We are, so far, not aware of any direct evidence for this self-similar character of the renormalized ion concentration, $c^{*}$, although the observed linearity of the screening length with the concentration suggests such a picture. To support this argument, we presented a physical argument for the electrostatic energy of an unpaired ion which identifies the region $R$ as the volume around an unpaired ion with a higher dielectric constant than the average.  The self-energy of such an ion includes this local effect \cite{RN79, RN80} as well as the average potential in the system due to screening, and the total electrostatic energy of the ion is minimized when $R\sim 1/\kappa$, providing more physical evidence for our second scaling argument.  In addition, the physical argument indicates why the screening length, while proportional to $\ell \, cd^3$ is numerically larger than the Bjerrum length even though the factor $cd^3$ is smaller than one. This is due to an additional small parameter, the difference in the local dielectric constant near the unpaired ion and the system-wide average of the dielectric constant.  This,  small parameter enters the denominator of the screening length given Eq.~\ref{finallambda}, allowing the screening length to be relatively large.   We note, moreover, that all of our arguments are  coarse grained and rely on the screening length being large enough that the surroundings of an unpaired ion can be described by a local dielectric constant which has contributions from both the clusters and the water in the vicinity of the unpaired ion.  This suggests that the Perkin screening law may only be observable in experiments that probe the system on large enough scales (and at long enough times).  Perhaps this is related to the fact that AFM experiments \cite{RN143} that probe the electrostatic forces at relatively small scales, as well as some computer simulations do not see evidence for long-range underscreening.  

Finally, we note that similar to the limits on the DH theory at small salt concentrations, our renormalized screening length description is only valid when the screening length is larger than the distance between unpaired ions: $\lambda > 1/{c^{*}}^{1/3}$.  This is equivalent to requiring that the volume fraction of ions still be relative large, $c d^3 > \delta^{3/2}$, but for small enough $\delta$ this is not a serious limitation since the experiments and theory presented here apply to the regime of relatively large salt concentration.

\section*{Conflicts of interest}
There are no conflicts to declare

\section*{Acknowledgements}
We thank Susan Perkin for useful discussions as well as F. Mugele, D. Andelman, H. Diamant, R. Adar, V. Craig, A. Ciach, C. Holm, S. Kathmann, B. Rotenberg, S. Marcelja, G. Fredrickson, and Bohan Chu.  We are grateful to P. Grassberger for the picture of clusters and the Perlman Family Foundation for their historic support.  

\bibliography{Electrostatics.bib}

\end{document}